\documentclass{article}

    \PassOptionsToPackage{numbers, compress}{natbib}
\usepackage[final]{neurips_2019_ml4ps}

\usepackage[utf8]{inputenc} % allow utf-8 input
\usepackage[T1]{fontenc}    % use 8-bit T1 fonts
\usepackage{hyperref}       % hyperlinks
\usepackage{url}            % simple URL typesetting
\usepackage{booktabs}       % professional-quality tables
\usepackage{amsfonts}       % blackboard math symbols
\usepackage{nicefrac}       % compact symbols for 1/2, etc.
\usepackage{microtype}      % microtypography
\usepackage{graphicx}
\graphicspath{ {./figures/} }

\usepackage{wrapfig}

\title{On the loss of learning capability inside an arrangement of neural networks}

\author{%
   Ivan Arraut \\
   The Open University of Hong Kong \\ 
   30 Good Shepherd St., Kowloon, Hong Kong, China\\
  \texttt{ivanarraut05@gmail.com} 
  \And
   Diana Diaz \\
   Wayne State University \\
   5057 Woodward Ave., Detroit, MI, USA \\
   \texttt{dmd@wayne.edu} 
}

\begin{document}

\maketitle

\begin{abstract}
We analyze the loss of information and the loss of learning capability inside an arrangement of neural networks. Our method is new and based on the formulation of non-unitary Bogoliubov transformations in order to connect the information between different points of the arrangement. This can be done after expanding the activation function in a Fourier series and then assuming that its information is stored inside a Quantum scalar field.
\end{abstract}

\section{Introduction}

Artificial neural networks were proposed for the first time in 1943 as an attempt to simulate the way how the human brain operates \cite{1}. By then, the logic, taken as the way how some input information is interpreted, was the key ingredient for the formulation of this concept. Subsequently, the notions of Hebbian learning were developed \cite{2}. The Hebbian network was after analyzed in \cite{111}. The perceptron was proposed in 1958 by Rosenblatt by assuming that there exists a bridge between psychology and biophysics. The perceptron is based on three fundamental questions: 1. How is information about the physical world sensed or detected, by the biological system? 2. In what form is information stored or remembered? 3. How does information contained in storage or memory, influence recognition and behavior?.\\ By complementing the perceptrons with connections (synapses in biological terms), and by taking the output of a neuron as the input of another one, we create a primitive arrangement of neural networks. Although this is a good starting point, it is difficult for an arrangement of neural networks based on perceptrons to learn in the sense of Machine Learning. The reason is that when we use perceptrons, a small change in the input can create huge changes in the output, destroying in many cases the possibility of learning \cite{MN}. When a system of neural networks learns, normally it makes improvements based on variations on the weights (importance of the information transmitted through the synapses) and the bias (related to the threshold). Examples of this can be found in the implementation of the method of {\it gradient descend} to minimize the cost-function \cite{grad}. In such a  case, the perceptrons would operate terribly because any attempt of learning some specific output information from a single output neuron, might destroy easily what has been learned in other output neurons. This situation can be solved by introducing an activation function $f(x)$, different to the standard step function in this scenario. The most common non-trivial activation function used is the sigmoid function and here we will take it as the ideal output of a single neuron. The biggest advantage of the sigmoid function is its high sensitivity in responding to small changes in the inputs. This means that it can map the small input changes into small output changes. Those small changes might be entered through variations of the weights and bias. In this way, the neural network will be able to learn by updating these values until it gets the desired result. Normally the process of learning is based on the minimization of the cost function. Once the system learns something, ideally it does not forget it and the information is in principle preserved. However, there are cases where there might be a loss of information due to different effects and this might affect the capability of the system for learning.\\ In this paper, we introduce a novel method for analyzing the loss of learning capability and/or loss of information inside a neural network arrangement. Our method is based in the Fourier expansion of the activation functions, taken as scalar Quantum fields \cite{sq}. We then employ the concepts of {\bf Bogoliubov transformations} \cite{Bogo} to relate the information going from one neuron to another one as it was proposed in \cite{Ivan}. In this scenario, we interpret the loss of information as the loss of unitarity in the transformation. In this way, we evaluate the conditions under which the loss of unitarity transforms our sigmoid response functions (a system able to learn) into a standard step function (system unable to learn), destroying then the capability of the network to learn. Interestingly, we understand that the amount of information stored is also connected with the learning capability of the system, just as it happens in biological systems. The paper is organized as follows: In Sec. (\ref{s1}), we introduce the basic concept of perceptrons, understanding that the activation function is a step function in such a case. In Sec. (\ref{s2}), we explain the neural network arrangement using the sigmoid as an activation function. In Sec. (\ref{s3}), we analyze the evolution of the information in the system, including loses, by storing the information locally inside a scalar Quantum field. We then analyze how much of this information is lost, after crossing through the synapse toward another arbitrarily selected neuron. This is achieved by using the famous Bogoliubov transformations \cite{Bogo}. Finally, in Sec. (\ref{conc}), we conclude.
\section{Perceptrons: Basic concepts}   \label{s1}
\begin{wrapfigure}{r}{0.31\textwidth}
    \includegraphics[width=0.33\textwidth]{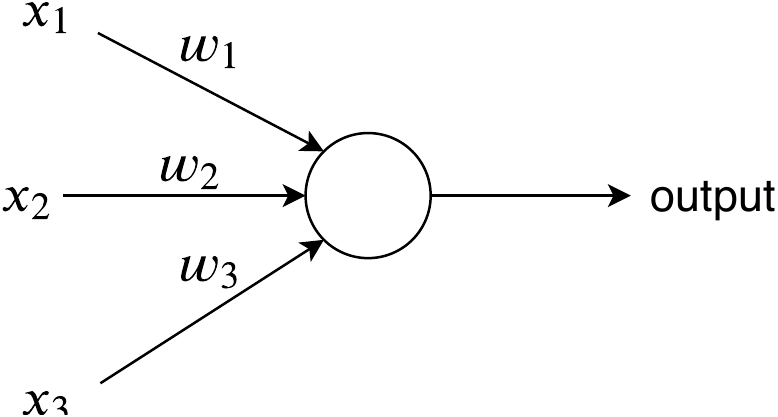}
  \caption{Perceptron neuron with three input variables with a  single output $0$ or $1$. The inputs are represented by $x_1$, $x_2$ and $x_3$. Here $\omega_i$ are the weights corresponding to each input.}
  \vspace{-5pt}
  \label{F1}
\end{wrapfigure}
Let's start with the basic definition of neural networks, by considering each neuron to be equivalent to a perceptron in the form originally proposed in \cite{1}. A perceptron takes different inputs defined by a set of variables $x_1$, $x_2$, $x_3$,... $x_n$ and it reproduces a single binary output \cite{MN}. The neuron in such a case reproduces some specific output $m_j$ defined by either, $m_j=0$ or $m_j=1$ ($j=1, 2, 3,...$ are the number of output neurons), depending on whether or not the weighted sum of inputs is larger or not than some specified threshold. The threshold can also appear in the form of bias. Specifically, the condition imposed over the perceptrons is
\begin{eqnarray}   \label{defined}
A=0,\;\;\;if\;\;\;\sum_j\omega_jx_j+b\leq0\nonumber\\
\;=1,\;\;\;if\;\;\;\sum_j\omega_jx_j+b>0.
\end{eqnarray}
Here $b$ corresponds to the bias of the perceptron and it is equivalent to the threshold of the neuron. Figure (\ref{F1}) illustrates the standard behavior of a perceptron neuron.\\ We can perceive the perceptron as a neuron able to make a decision based on input weighted evidence. From the perspective of biological neurons, the weights $w_l$ defined in eq. (\ref{defined}) represent the importance which the synapse gives to the different input patterns. One example with two weights can be the decision of wearing a certain color of clothes. We can assume that the output $A=0$ means dressing in black or blue color and the output $A=1$ means dressing any color different to black or blue. We can then take the first binary input as $x_1=1$ if there is at least one protest in the streets, and $x_1=0$ if there are no protests. Similarly, we might take the second input as $x_2=1$ if the protesters dress black and the policemen dress blue (or vice-versa); and $x_2=0$ if the protesters and the policemen dress any color different to black or blue. Based on these, our system has to decide if it is good to dress black or blue during a protest season as the one lived in Hong Kong. Our system could give weights to the pair of binary inputs, let's take the same weights for both inputs as $w_1=2$ and $w_2=2$. This means that both inputs are equally important for the system. If we take the bias to be equivalent to $b=-3$, then in agreement with eq. (\ref{defined}), the system will decide to dress black or blue only if there are no protests on the streets or if the protesters and police (in case of having protests on the streets) are not dressing the mentioned colors (black or blue). On the other hand, the decision of not dressing the same colors is only taken if both situations appear, namely, there is a protest {\bf and} the police and protesters are dressing in the mentioned colors. In this simple situation, a system with an operation based on perceptrons will be enough. More complex arrangements can be done if we create a more complex network like the one in Figure (\ref{F2}). To show the difficulties that the perceptron neurons have in learning appropriately some specific tasks, we study sigmoid neurons and focus on this important aspect of artificial neural networks.
\section{Sigmoid neurons}   \label{s2}
The sigmoid neurons are different from the perceptrons because the response function  of the neurons is not a step function, but rather a sigmoid function defined as
\begin{equation}
\sigma(z)=\frac{1}{1+e^{-z}},    
\end{equation}
where $z=\sum_j\omega_jx_j+b$, which is the same entrance used for the perceptrons in eq. (\ref{defined}). Note that the response function can take any real value between $0$ and $1$ which marks a significant difference from the perceptron case. This corresponds to a huge advantage for the process of learning when compared with the perceptron. Note that there are two limits for the sigmoid function where the behavior corresponds to perceptrons. This happens when $z=\pm\infty$. The biggest advantage of the sigmoid function for learning is that small changes in the inputs correspond to small changes in the outputs. This can be quantified mathematically as
\begin{equation}   \label{This one}
\Delta m\approx \sum_j \frac{\partial m}{\partial \omega_j}\Delta\omega_j+\frac{\partial m}{\partial b}\Delta b. \end{equation}
This is only the ordinary chain rule in Calculus. The superiority in learning for neural networks using as a response to the sigmoid function can be seen in the example of
a simple network to classify handwritten digits
using a three-layer neural network.
%where the system tries to identify a handwritten number from $0$ to $9$. Such an arrangement can be seen in Figure (\ref{figure7}), 
The input layer of the network contains neurons encoding the values of the input pixels and the output contains $10$ neurons, each one representing a different digit from $0-9$. If the first neuron is excited, then the system will identify $0$, if the second output neuron is the one excited, then the system will naturally identify $1$ and so on. Let's assume that the system has a problem identifying the number $9$, i.e. the output neuron, corresponding to the digit $9$, does not get excited even if the input image is a $9$.
%This would mean that even if the input image is the number $9$, the last output neuron, corresponding to the same number, does not get excited.
Then, the system would have to be adjusted by changing gradually the weights and bias, such that the last neuron can shoot for the appropriate case. This is what we call the process of learning. The best way for doing this is to make small changes in the weights and on the bias until the system can identify the number $9$. The small changes in weights and bias will be detected as small changes in the outputs in agreement with eq. (\ref{This one}). This gives then the advantage of testing simultaneously all the other numbers. Or equivalently, in this way we avoid to damage any adjustment for the other output neurons corresponding to the identification of the other numbers. This clever way of learning is impossible if we use perceptrons because for perceptrons we would have possible drastic changes in the outputs for a finite change in the inputs. It is for this reason that the possible transitions from the sigmoid response to the perceptron response, as a consequence of the loss of information, deserves careful attention.
%
%\section{The loss of information in neural networks and its connection with the loss in learning capability}   \label{s3}
\section{Information loss in neural networks and loss in learning capability}   \label{s3}
\begin{wrapfigure}{r}{0.5\textwidth}
  \begin{center}
     \includegraphics[width=\linewidth]{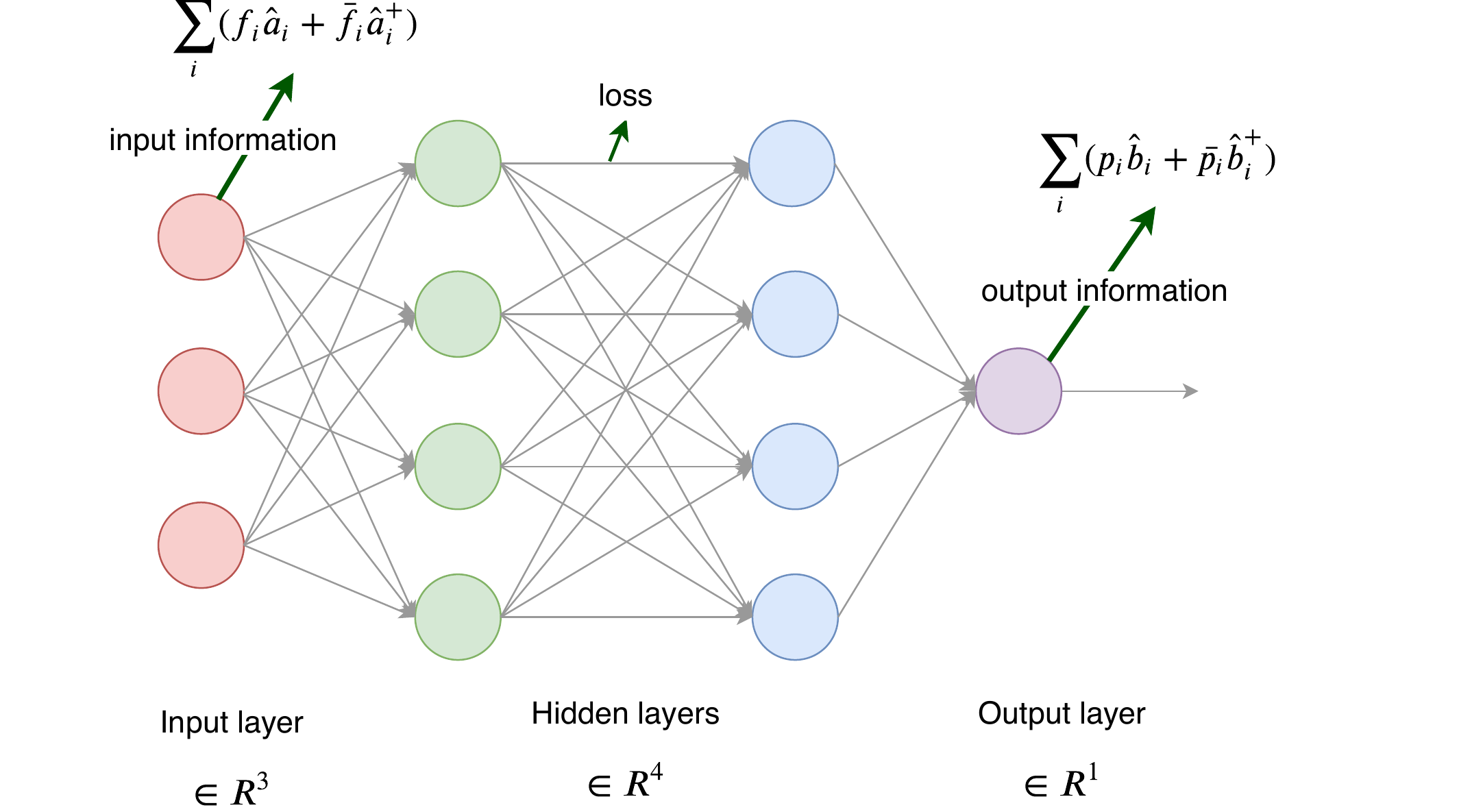}
  \end{center}
    \caption{Standard neural networks. The information flows from the input to the output. There might be lost of information during the transmission through the synapses as the figure illustrates. We take the input information as it is stored in a Quantum field. The same occurs for the output information. The loss of information is reduced to zero if $[\hat{b}_k, \hat{b}^+_{k'}]=\delta_{k, k'}=[\hat{a}_k, \hat{a}^+_{k'}]$. The information is loss for the output when $[\hat{b}_k, \hat{b}^+_{k'}]\neq[\hat{a}_k, \hat{a}^+_{k'}]$.}
    \label{F2}
\end{wrapfigure}

Intuitively it is expected (although not obvious) that the loss of information in a system must be related to the loss of the ability of the system for learning. 
%Here we extend the method proposed in \cite{Ivan}.
Here we use the method proposed in \cite{Ivan} and extend it to real scenarios.
First, we redefine the functions as Quantum scalar fields storing some amount of information. In this way, we define the Quantum field for the Heaviside function as
\begin{equation}   \label{distr}
<0\vert\hat{\phi}_1(z)\vert0>=f_p(z)=\sum_{k=-\infty}^z\delta[k]=<0\vert\sum_k\left(p_k(z)\hat{b}_k+\bar{p}_k(z)\hat{b}^+_k\right)\vert0>.
\end{equation}
Here $\delta[k]$ is the standard delta-Dirac function which characterizes the step function (perceptron) behavior and $\vert0>$ corresponds to the standard vacuum state. Here however, the vacuum $\vert0>$ is not the vacuum for the modes $p_k(z)$ as we will explain in a moment. In eq. (\ref{distr}), the right-hand side defines the function as a field expanded in terms of positive and negative frequency modes $p_k(z)$ and $\bar{p}_k(z)$, with the corresponding annihilation and creation operators $\hat{b}_k$ and $\hat{b}^+_k$ respectively. Figure (\ref{F2}) shows a point in the network where the information stored in the system obeys the distribution (\ref{distr}). Note that we are taking the step function behavior as the total output of the system. Locally, the operators obey some local algebra
\begin{eqnarray}  
[\hat{b}_k, \hat{b}^+_{k'}]=\delta_{k, k'},\;\;\;\;\;information\;preserved,\nonumber\\
\neq\delta_{k, k'}\;\;\;\;\; information\;lost.
\end{eqnarray}
For the output neuronal response function, we define a vacuum state as $\hat{b}_k\vert\bar{0}>=0$.\\
In the same way, we can expand the sigmoid function by identifying it with the scalar field
\begin{equation}   \label{QF1}
<0\vert\hat{\phi}_2(z)\vert0>=f_\sigma(z)=\frac{1}{1+e^{-z}}=<0\vert\sum_k\left(f_k(z)\hat{a}_k+\bar{f}_k(z)\hat{a}^+_k\right)\vert0>.   
\end{equation}
In a standard form, we will take the algebra of creation and annihilation operators for this field as fixed and being equivalent to
\begin{equation}
[\hat{a}_k, \hat{a}^+_{k'}]=\delta_{k, k'}.    
\end{equation}
In Figure (\ref{F2}), we can see that the sigmoid information appears complete on the input neurons. This means that we are taking the Quantum field (\ref{QF1}) as a field containing all the information. There are no loses at this point. Here we also define another local vacuum condition $\hat{a}_k\vert0>=0$. Note that this vacuum $\vert0>$ is different to the vacuum $\vert\bar{0}>$ defined for the $\hat{b}_k$-operators. This means that in different points of the network we have a different amount of information, and at each point, we define "empty" or "no-information" in a different way. We can then connect the operators defined in eq. (\ref{distr}) with those defined in eq. (\ref{QF1}) via Bogoliubov transformations. Then we have the relations
\begin{equation}   \label{beta}
\hat{b}_k=\sum_k\left(\bar{\alpha}_{kj}\hat{a}-\bar{\beta}_{kj}\hat{a}^+_j\right).    
\end{equation}
For our purposes, this is the important relation because it is the one which establishes the connection between the local vacuums defined at each point of the arrangement. Note that if $\beta_{ij}=0$ in the previous equation, then the information is preserved and $\vert0>=\vert\bar{0}>$ unambig\"uously. This is the case because in these special circumstances, both operators $\hat{a}_k$ and $\hat{b}_k$ annihilate the same vacuum. Then in these special circumstances $[\hat{a}_k, \hat{a}^+_{k'}]=\delta_{k, k'}=[\hat{b}_k, \hat{b}^+_{k'}]$. For this case the sigmoid response function will never change into a step function. On the other hand, if $\bar{\beta}_{kj}\neq0$, then we have loss of information and then $[\hat{a}_k, \hat{a}^+_{k'}]=\delta_{k, k'}\neq[\hat{b}_k, \hat{b}^+_{k'}]$. If we want to search for the information loss, we can expand the sigmoid function (\ref{QF1}) in terms of the modes of the step function (\ref{distr}) as
\begin{equation}   \label{QF12}
<0\vert\hat{\phi}_2(z)\vert0>=f_\sigma(z)=\frac{1}{1+e^{-z}}=<0\vert\sum_k\left(p_k(z)\hat{b}_k+\bar{p}_k(z)\hat{b}^+_k+q_k(z)\hat{c}_k+\bar{q}_k(z)\hat{c}^+_k\right)\vert0>.   
\end{equation}
the above result suggests that to reproduce the information of the sigmoid function using as a starting point the step response function, it is necessary to add some extra modes $q_k$ with their corresponding operators $\hat{c}_k$ to the step function. Then it is possible to conclude that the modes leaving the system and representing the loss of information are
\begin{eqnarray}   \label{QF12again}
 <0\vert\hat{\phi}_2(z)-\hat{\phi}_1(z)\vert0>=f_\sigma(z)-f_p(z)=<0\vert\sum_k\left(q_k(z)\hat{c}_k+\bar{q}_k(z)\hat{c}^+_k\right)\vert0>\\
 =\frac{1}{2}\left(tanh\left(z/2\right)+1\right)\nonumber -\frac{1}{2\pi sinh\left(z/2\right)}\int_{-\infty}^\infty dz(cos(kz)-isin(kz)).
\end{eqnarray}
This expansion represents the information escaping the system when it flows from the input toward outputs. The conditions for getting a step function from an initial input sigmoid are then reduced to the analysis of the matrix elements $\bar{\beta}_{kj}$ in eq. (\ref{beta}). Note that the difference between the equations (\ref{QF12}) and (\ref{QF12again}) is the presence of the the modes expanded by the operators $\hat{b}_k$, which are the modes perceived by the system after losing information. The $\hat{b}_k$-modes can be quantified in a similar way as Hawking quantified the modes escaping the event horizon of a Black-Hole \cite{Hawking}. Then our work, besides contributing to the understanding the loss of learning capability of a neural network arrangement, it also provides the opportunity of testing some possible scenarios where we can find analogue models where the Hawking radiation and similar effects could be tested. As a final remark, note that all the vacuum expectations values are evaluated with respect to a common vacuum $\vert0>$. This corresponds to the vacuum for the modes $f_k(z)$. Only if we use a common vacuum, we can compare the amount of information stored at different locations of the network. This technique is original and it is the main contribution of our paper.     
%%%%%%%%%%
\section{Conclusions}   \label{conc}
Here, we have presented a novel method for analyzing the loss of information in a neural network. The method also connects the loss of information with the loss of the system's learning capability. It consists of promoting the response functions as Quantum fields expanded as a function of positive and negative frequencies. If the information is preserved in the system, the Bogoliubov coefficient $\beta_{kj}$ would vanish and then the algebra of commutators would be also preserved. This in Quantum mechanics is known as unitarity. On the other hand, a non-trivial value of $\beta_{kj}$ would imply the non-conservation of information and then the possibility of losing the advantages of the sigmoid response function. Similar effect appears in the evaporation of a Black-Hole and it is known as Hawking radiation \cite{Hawking}. This effect can be cumulative and the system effectively could behave as a standard perceptron even if the neurons were configured originally with a sigmoid response function. The take-home message in this work is that if we want to quantify the loss of information in a system storing information in agreement with a bosonic field, this can be quantified by a non-unitary Bogoliubov transformation. The non-unitarity registers the amount of information that we can never recover again and in addition affects our system in its learning process. Note that the loss of information explored here is different to the information in the way analyzed in \cite{Final1, Final2}. In fact, the extraction of information based in the Bottleneck principle promoted in \cite{Final1, Final2} consists in the elimination of what we interpret as an initial noise. We could use similar techniques to the one showed in this paper for explaining such effect, but we will let this part for a future paper. Our paper instead deals with the loss of useful information, namely, the information which disappears assuming that we are able to eliminate the noise in advance. More details about this will be also given in a more formal paper. In subsequent works, we will also deliver more details of our analysis, including possible algorithms, which will take time to get them ready but are under process. The limitations in this task are connected to the limitations in the understanding of the storage of information by using Bosonic systems. Bosonic systems are able to store more information than qubits but they can also lose it easier. This means that our analysis of loss of information in this paper is extremely important for the development of more efficient systems for storing information.      

%\begin{figure}[h]  
    %\includegraphics[width=.6\linewidth]{digits}
    %\caption{A neural network designed for identifying the digits from $0$ to $9$. Note that for this purpose the arrangement has $10$ output neurons. This case is the best example for understanding why the neurons using the sigmoid function as a response function are superior to the perceptrons for learning complex tasks.}
%\label{figure7}    
%\end{figure}

\small

\bibliographystyle{unsrt}  


\end{document}